\documentclass[iop]{emulateapj}

\shorttitle{Stability of P-type orbit}
\shortauthors{Chaelin Hong \& Maurice H.P.M. van Putten}

\begin{document}


\title{On the stability of corotating and counter rotating P-type orbits around stellar binaries: a numerical study}


\author{Chaelin Hong\altaffilmark{1}}
\affil{CHEA-SRC, School of Natural Science Ulsan National Institute of Science and Technology (UNIST)}

\author{Maurice H.P.M. van Putten\altaffilmark{2}}
\affil{Astronomy and Physics, University of Sejong}

\altaffiltext{1}{CHEA-SRC, School of Natural Science Ulsan National Institute of Science and Technology (UNIST),
\\ Eonyang-eup, Ulju-gun, Ulsan 44919, Republic of Korea\\email: chealin93@gmail.com}

\altaffiltext{2}{Eonyang-eup, Ulju-gun, Ulsan 44919, Republic of Korea\\
Astronomy and Physics, University of Sejong, 209 Neungdong-ro, Gwangjin-gu. Seoul 143-747, corresponding author,\\email: mvp@sejong.ac.kr}


\begin{abstract}
We here revisit the essential problem of dynamical stability of planetary orbits around stellar binaries. We build on the coplanar three-body system of the Dvorak(1986), extending his stability diagram to both corotation and counter rotation of P-type orbits. His stability diagram express the change of stability across the gap between upper(UCO) and lower(LCO) critical orbits. By radius, this gap has a width of about 8$\%$ in the corotation case and 24$\%$ in the counter rotation case. As the gap of the second lies below the first, counter rotation is more stable, yet by width it is more chaotic. The gap between UCO and LCO follows a transition radius $r^+_g = 2.39 + 2.53e - 1.40 e^2$ and $r^-_g  = 0.92 - 2.47e$ for corotation and respectively, counter rotation of the third body (the planet). Our $r^+_g$ agrees with the same of Dvorak to within 0.35$\%$. As a result, we discover $r^+_g/r^-_g \lesssim 2.57$ for all $e$. Around dim binaries, therefore, a relatively close in habitable zone may still be populated with planets on counter rotating orbits.
The accurate numerical results presented here based on adaptive integration using MATLAB ODE45 may also serve as a novel benchmark of accurate \textit{N}-body integrators of exosolar systems more generally.
\end{abstract}





\section{Introduction}

A new field of searches for exoplanets and exoplanetary systems is emerging since the discovery of the first extrasolar-planet orbiting a main-sequence star 51 Pegb \citep{Mayor1995}. Various exoplanet databases are compiled by several group to provide the scientific community for improving our understanding of planet with a host star based on physical properties \citep{DolevBashi2018}, that continue to grow by, e.g., the recently launched Transiting Exoplanet Survey Satellite (TESS). Exploration of detailed physical properties such as atmospheric composition \citep{kem18} and planetary spin \citep{sne14} of these exoplanets in their respective habitable zone \citep{kas93,kop13} will be possible by the upcoming advanced James Webb Space Telescope (JWST) and the Extremely Large Telescope (ELT). 

Interestingly, a fair number, though small in percentage, of exoplanets are discovered about double stars systems. For these exoplanets, the habitable zone must include dynamical stability of planetary orbits. A common classification of these three-body systems is in terms of P-type and S-type planetary orbits, where the first is circumbinary and second is around either one of the primaries, and L-type is when the planet librates around one of the triangular Lagrangian points. The first detection of such is the S-type planetary orbit in the Kepler system \citep{egg04,egg07,des07,roe12}, followed by the P-type orbit of Kepler-16b \citep{doy11}.

For the P-type orbit, a concise stability diagram was first introduced by \citep{Dvorak1986} for the coplanar plane as a function of eccentricity of the central binary. Various extensions have been considered, e.g. finite mass ratios when planetary masses are small \citep{Holman1999,Cuntz2015}, and theoretical studies on the orbital resonances \citep{Morais2012}.

While to observers, the inclination angle of planetary orbits to that of the central binary relevant, for orbital stability, the effect on planetary orbital stability is rather weak \citep{Pilat2002}. 

For potential habitability for advanced life, the exoplanet should have low spin to permit a global clement climate to emerge \citep{vanputten2017}. If exoplanets are born with arbitrary spin, this probably requires de-spining by lunar tidal interactions rather than stellar tidal interactions at canonical distances of the habitable zone. Exomoon therefore may serve a proxi for potentially advanced life on exoplanets in their respective habitable zone. Earth, for instance, has spun down to 24 hour/day from 4.1 hours/day initially by lunar tidal interaction over the past 4.5 Gyr. Early on, relatively strong tidal interactions may have been relevant to biotic processes and abiogenesis \citep{Lingam2017}, while this would be disadvantagous to development of a global clement climate conducive to advanced life.

Selecting exoplanets for follow-up observations, therefore, involves detailed consideration of orbital stability of exoplanet - moon systems. To assess their stability, we build an accurate exosolar \textit{N}-body simulator that can handle a large mass hierarchy defined by mass ratios of exoplanet to central binary \textit{and} exomoon to exoplanet.

In this report, our starting point is a revisit and extension of the work on the stability of P-type planetary orbits pioneered by \citep{Dvorak1986}. Specifically we extend his analysis of the three-body problem to include the unrestricted three-body problem and counter-rotating orbits, all in coplanar configuration.  

For P-type orbit, Dvorak showed that a change of stability occurs across a gap between stable and unstable regions outside \textit{Upper Critical Oribit} (UCO) and inside the \textit{Lower Critical Orbit} (LCO). UCO and LCO refers to the initial orbital radius of the third body, whose orbits are stable and, respectively, unstable after integration time over a fixed number of periods of the central binary. We study different behavior of UCO and LCO for the co- and counter rotation case. Our numerical results confirm and extend the stability diagram of Dvorak and are presented here also to serve as potential benchmarks for \textit{N}-body simulators more generally. To this end, we pay specific attention to the problem of numerical convergence as a function of the number of orbital periods. 

Our exosolar system \textit{N}-body simulator employs the ODE45 solver of MATLAB \citep{MathWork}. This is a fourth and fifth order adaptive ordinary differential equation solver. ODE45 performs accurate integration with numerical errors down to numerical round-off error in double precision of total-energy in the problem setting at hand, that focuses on a limited integration time. For very long integration time, \citep{Zeebe2017} reviews many different methods for accurate integration that are outside the scope of our study.  
As a function of eccentricities from 0 to 0.9, We compute our extended stability diagram according to the following:  

\begin{enumerate}
\item \textit{Restricted and unrestricted three-body problem.}
 In the three-body problem, the force of mutual attraction between the primaries and the third body is proportional to the mass of the latter. As the third body is taken to be small mass, it does not significantly influence the primaries. This system is called the restricted three-body problem. Stability is slightly perturbed in the un-restricted problem with small masses, that is readily resolved numerically at high resolution. 

\item \textit{Corotation and counter rotation.}
 Since stability of P-type planetary orbits is expected to be sensitive to the angular velocity relative to the angular velocity of the central binary, corotation and counter rotation are expected to give different UCO and LCO, here studied side-by-side with otherwise the same set-up. In fact, since the angular velocity difference between that of the planet and the primary is relatively greater in the second case, we anticipate this second case to be relatively more stable.

\item \textit{Convergence} of our numerical results against variation of integration time. We confirm that few times period orbits of the binary is moderate to integration time by probing convergence the same UCO and LCO as a function of orbital integration time. 

\end{enumerate}

Our roadmap is as follows. In \S2, we discuss an initial set-up of restricted three-body problem including finite mass ratios of third body to study the stability of P-type orbits.  
\S3, details the dynamical equations of motion. Results of P-type orbits stability are studied in \S4. In \S5, we discuss our results. 

\section{P-type orbits : initial conditions}

To study dynamical stability of P-type orbits in the coplanar three-body problem, we set initial positions of three objects with dimensionless variables normalized to the solar system unit of length (1AU) and solar mass (\(M_\odot\) = $2\times10^{33}$\,g). The center of mass (CM) is at the origin (0,0,0) of Cartesian coordinate system. Our equal mass primary binary has initial separation 1. The mass of the third body, $m_3$ is $10^{-8}$ of the total mass of the binary, $M = m_1 + m_2$. 

\begin{figure}
   \centering
  \includegraphics[width=0.5\textwidth,height=5cm]{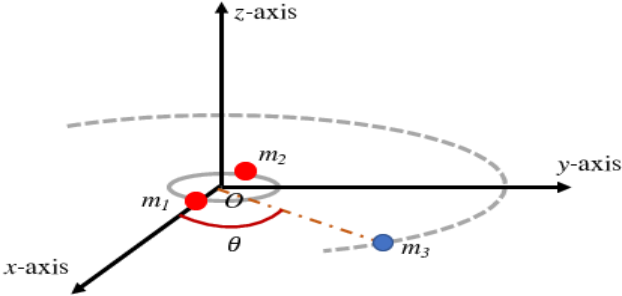}
      \caption{ Configuration of initial data. All three bodies have a coplanar configuration. The center of mass (CM) of the two primaries is at the origin (0,0,0) in a Cartesian coordinate frame $(x,y,z)$. 
      Initial distances of the third body range from 0.55 to 4.3 AU from the CM with azimuthal $\theta$ to the line connecting the two primaries.}
         \label{Fig1}
   \end{figure}

In this configuration, closely following Dvorak (1986), our initial distance between the CM and the third body is varied from 0.55 up to 4.3 (step size 0.05). A second variable in the initial data is the angle $\theta$ defined by the angle to the line connecting the primaries (Fig.~\ref{Fig1}, Table~\ref{table1}). This initial angle is varied from 0 to 180$^\circ$ (step size 45$^\circ$). We integrate the three body motion over $N$ Dvorak's orbital study with $N=500$ up to $N = 16000$ periods of the central binary. The primaries with initial eccentricity, $e$, have an initial circular velocity about the CM. In all experiments, the third body is also given an initial circular velocity about the CM. 

Upon integration over $N$ periods of the central binary, first, we identify initial distance and $\theta$ which gives rise to a stable or unstable orbit. If for all $\theta$, the given initial distance ensures stability, then this distance belongs to the region of stability which is bounded below by the UCO. On the other hand, if it is unstable, it belongs to the region of instability which is bounded above by the LCO. Repeating this procedure obtains UCO and LCO as a function of aformentioned eccentricity. 
 
Table~\ref{table1} lists our parameters and those of Dvorak whose original work uses slightly different values. Following Dvorak, the four distinct initial angle $\theta$ measured a small sample of $( 0^\circ \le e < 180^\circ)$, that otherwise may contain an infinite number of $\theta$ values of interest to orbital stability given the chaotic nature of the three-body problem.   
   \begin{table}
      \caption[]{Initial data in the P-type stability problem where $e$ refers to eccentricity of the central binary. Here, $e$ refers to initial eccentricity in the unrestricted problem.}
         \label{table1}
     $$ 
         \centering
         \begin{tabular}{llllcccc}
            \hline
             & This work & Discretization & Dvorak (1986) \\
            \hline\hline
            $\theta$  & $0 - 180^\circ $  & $45^\circ$ & $0 - 145^\circ$ \\
            $e$ & $0 - 0.9$  & 0.02 & $0 - 0.9$ \\
            $r$ & $0.55 - 4.3$ & $0.05$  & $0.8 - 4.1$ \\
            $N$ & $2^m$ & $m=9,...,14$  & $500$ \\
            \hline
         \end{tabular}
     $$ 
   \end{table}
\section{Dynamical equations of motion}

The general three-body problem for masses $m_i$ with position ${\bf x}_i$ ($i = 1,2,3$) and Newton's constant $G$ is described by the following the dynamical equations of motion in time $t$:
 \begin{equation}
\begin{array}{lll}
\ddot{{\bf x}}_1 &=& - Gm_2 \frac{  {\bf x}_1 - {\bf x}_2  }{||{\bf x}_1 - {\bf x}_2 ||^{3}}  - Gm_3 \frac{  {\bf x}_1 - {\bf x}_3  }{||{\bf x}_1 - {\bf x}_3 ||^{3}},\\\\
 	\ddot{{\bf x}}_2 &=& - Gm_3 \frac{  {\bf x}_2 - {\bf x}_3  }{||{\bf x}_2 - {\bf x}_3 ||^{3}}  - Gm_1 \frac{  {\bf x}_2 - {\bf x}_1  }{||{\bf x}_2 - {\bf x}_1 ||^{3}},\\\\
 	\ddot{{\bf x}}_3 &=& - Gm_1 \frac{  {\bf x}_3 - {\bf x}_1  }{||{\bf x}_3 - {\bf x}_1 ||^{3}}  - Gm_2 \frac{  {\bf x}_3 - {\bf x}_2  }{||{\bf x}_3 - {\bf x}_2 ||^{3}}.
\end{array}
\label{E1}
 \end{equation}

 In our numerical implementation, we use dimensionless variables, i.e., masses are normalized to 1 solar mass and distances to 1 AU, as mentioned before, and $G = 1$. 
As such, time $t$ is normalized to 1 yr. 

As a reference to integrating (Eqn.~\ref{E1}), we consider the total energy $H_1$ and eccentricity $e$ and semi-major axies $a$ of the $two$-body problem defined by central binary in light of exact solution:
\begin{figure}
   \centering
  \includegraphics[width=0.5\textwidth,height=7cm]{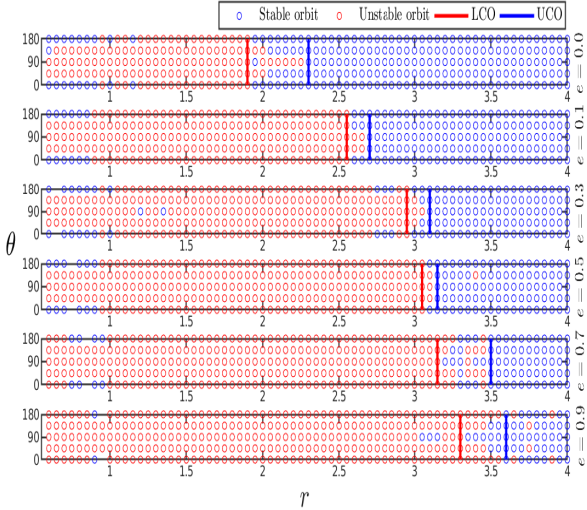}
      \caption{The change of stable (blue) to unstable (red) planetary orbits in plane of initial data ($\theta$, $e$) trends with ellipticity, $e$. 
      Noticeably, a few stable (unstable) orbits appear as islands in regions of otherwise unstable (stable) orbits, characteristic of this overall chaotic system ($N=512$).}
         \label{Fig2}
   \end{figure}  
\begin{equation}
\begin{array}{lll}
 H_1 &=& \mu \left( \frac{1}{2} v^{2} - GMu\right) = - \frac{GM\mu}{2a} ,\\\\
 u &=& \frac{GM}{j^{2}} \left( 1 + e\cos\varphi \right),
\end{array}
\label{E2}
 \end{equation}

 where $u = 1/r$ in a polar coordinate system centered about $M$, e.g., \citep{L.D.1969}.
Here, $H_1$ is the equivalent one-body hamiltonian wherein $\mu =  m_1m_2/(m_1+m_2)$ is the reduced mass, $v$ is the magnitude of velocity difference, $M$ is the total mass and $j$ is the specific angular momentum. 
By this analytic solution, we verifiy numerical errors in integration by ODE45 to be typically e.g.below $10^{-12}$, consistent with the relative and absolute error options set at that value. 

For the $three$-body problem (Eqn.~\ref{E1}), we split total energy $H = H_1 + H_2$ into the total energy (Eqn.~\ref{E2}) of the binary $H_1$ and the total energy of the third body, $H_2$, where $H$ is defined by the initial data. Whereas $H_1$ is exactly conserved in the restricted three-body problem, $H$ is exactly conserved in the unrestricted three-body problem. Thus, for the unrestricted problem, variations in $H_1$ will vary with the mass of the third body and linearly so when the mass of the third body is very small. For instance, doubling the mass of third body to 2$\times10^{-8}$ from 1$\times10^{-8}$ causes variations in $H_1$ to double. This is indeed observed in our numerical experiment. When the mass increases, we also confirm that the results remain essentially unchanged for the third body mass less than about $10^{-4}$. 

\begin{figure}
	\centering
	\includegraphics[width=0.5\textwidth,height=8cm]{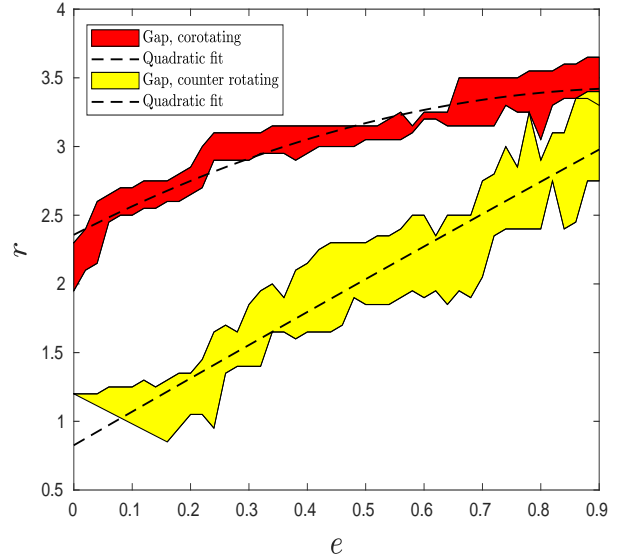}
	\caption{The gap between UCO and LCO as a function of $e$ for the corotating case (red) and counter rotating (yellow) and their quadratic fits, $r^\pm_g$, using polyfit function of MATLAB (black).
	 These two gaps refer to the unrestricted three-body problem with small mass planet. 
	 The width of gaps on average are 7$\%$ and 25$\%$ defined by the mean of UCO and LCO for corotation and counter rotation, respectively ($N=512$). Result for $N=8192$ are essentially the same, except for a slightly wider gap for corotation at $e=0.9$. 
	 }
	\label{Fig3}
\end{figure}
 We determine the UCO and LCO as a function of the initial $e$ of the central binary by a large number of runs following initial data in Table~\ref{table1}.
 
The three-body problem being inherently chaotic complicates the definition of UCO and LCO as boundaries of the regions of stability and, respectively, instability, since regions of stability (instability) inevitably contain isolated orbits that are unstable (stable) (Fig.~\ref{Fig2}). Nevertheless, the density of these isolated orbits is sufficiently low, that we can meaningfully define UCO and LCO, first demonstrated by Dvorak (1986), and here shown in Fig.~\ref{Fig2}, Dvorak (1986) further points out that stability of the third body is quite sensitive to the initial angle $\theta$ and $e$. the patter shows in $e$ general trend in UCO and LCO with $e$.


\section{P-type orbits: chaotic change of stability}

Fig.~\ref{Fig3} shows the overall increase of the UCO and LCO with $e$ for the data in Fig.~\ref{Fig2}, along with the same for our counter rotationg case. The \textit{gap} between the UCO and LCO is roughly uniform, apart from fluctuation inherent to chaotic behavior. The counter rotating case clearly lies below that of the corotating case: 
counter rotation is relatively \textit{more stable}, even though it is \textit{more chaotic} evidenced by a relatively wider gap (Fig.~\ref{Fig3}).

Specifically, the gap satisfies the following trends $( 0 \le e \le 0.9, N= 512)$:
\begin{equation}
\begin{array}{llll}
r^+_g  &=& 2.36 + 2.18e - 1.11e^2,\\\\
r^-_g &=& 0.83 - 2.45e - 0.07e^2.
\end{array}
\label{E3}
\end{equation}
\begin{figure}
   \centering
  \includegraphics[width=0.5\textwidth,height=8cm]{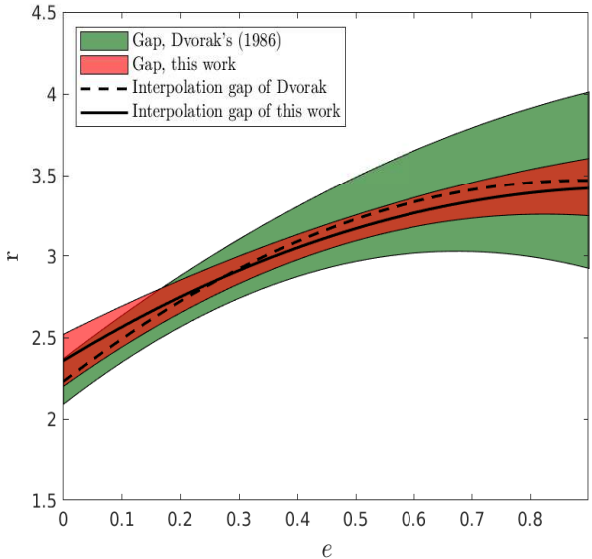}
      \caption{Comparison between the polyfit (black) of our results ($N=13$) and that (Dashed-black) of Dvorak's results from the gaps. The width of gap in our result is clearly narrow than the Dvorak's. }
         \label{FigVibStab}
         \label{fig4}
   \end{figure}
The associated mean widths of these gaps are 7$\%$ and, respectively, 25$\%$ for co- and counter rotation which is defined by the UCO and LCO($N=512$). And for $N=8192$ it is 7$\%$ and 24$\%$, the counter rotation is around three times narrower than the corotation. 
Our result (3) is essentially unchanged when we fall back to the restricted three-body problem in terms of the small mass of our planet. 

\begin{figure}
   \centering
  \includegraphics[width=0.5\textwidth,height=7cm]{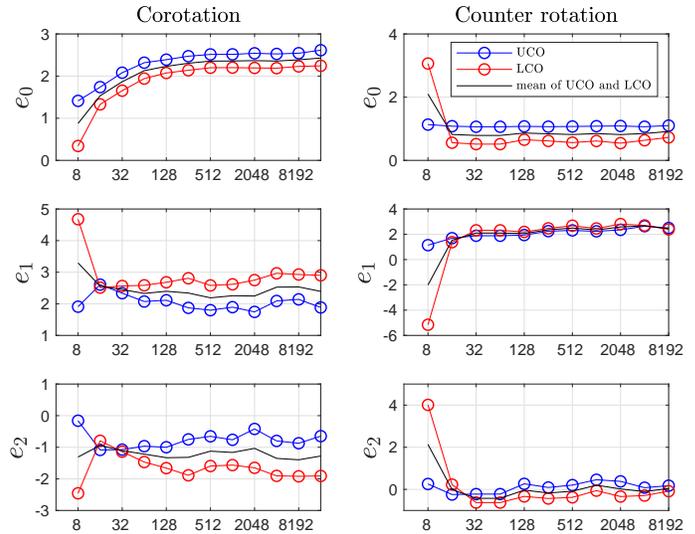}
      \caption{The convergence of coefficients of $r^+_g$ (left column) and $r^-_g$ (right column) for different $N$ integration time, $N$. In the each box, blue and red circles that are coefficients of the second order polynomial fitting of the UCO and LCO, respectively. For corotation, satisfactory convergence starts for $N \ge 512$ and for counter rotation, rapidly obtains that for $N \ge 16$.}
         \label{Fig4}
   \end{figure}

For the corotating case, Fig.~\ref{Fig4} shows a comparison of our trends $r^+_g$ with $r^+_D$, similarly defined for Dvorak's UCO and LCO, 
\begin{equation}
\begin{array}{lll}
r^+_D &=& 2.23 + 2.78e - 1.56e^2.\\\\ 
\end{array}
\label{EQN_dvorak}
 \end{equation}

The mean discrepancy between $r^+_g$ and $r^+_D$ is found to be a mere 0.35$\%$ even, counter.
On the other hand, our width of aforementioned 7$\% $($N=512$) is appreciably smaller than the width of 19$\%$ of the gap of Dvorak ($N=500$).

In other words, we agree on the overall trend in a change of stability in this inherently chaotic three-body problem, but find this change to more abrupt (our gap has smaller width). 

In order to check convergence, we extend our computation up to $N=8192$. We ascertain that convergence starts from around $N=128$, even though the nature of chaotic behavior remains evident in minor fluctuations in $e_1$, and $e_2$ as $N$ becomes large(Fig.~\ref{Fig4}). For $N=8192$, we report:

\begin{equation}
\begin{array}{llll}
r^+_g  &=& 2.39 + 2.53e - 1.40e^2,\\\\
r^-_g &=& 0.92 - 2.47e,
\end{array}
\label{E5}
\end{equation}
where the latter is effectively linear. 

\begin{figure}
   \centering
  \includegraphics[width=0.5\textwidth,height=7cm]{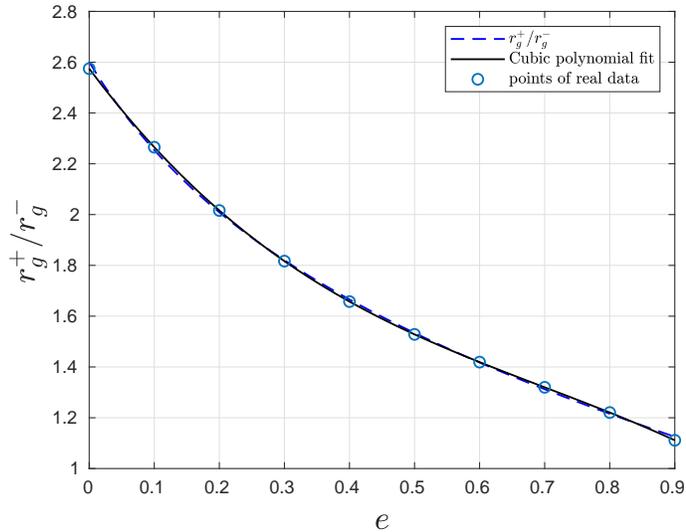}
      \caption{The ratio of $r^+_g$ to $r^-_g$. Data (circles, skyblue) are fit by cubic polynomial (solid, black).}
         \label{Fig6}
   \end{figure}  

\section{Conclusions}

We revisit the stability of planetary P-type orbits in the context of habitable zones around double star systems. The chaotic change of stability is identified in terms of the UCO and LCO originally introduced by Dvorak(1986). 
We extend this to the unrestricted three-body problem with corotation and counter rotation.
To this end, we use the ODE45 solver found to be a suitable integrator giving convergent results for integration periods extending up to $N=8192$. This is more than sufficient for our intended application creating synthetic transit light curves in future work. 

Our main finding is that counter rotation is appreciably more stable than corotation, which we attribute to a relatively higher difference in orbital angular velocity of the P-type planetary orbit relative to the angular velocity of the central binary. 

Relevant to the assessment of potentially habitable zones around double stars is the change of stability of P-type orbits. We find the gap between the UCO and LCO to be relatively narrow for corotation and somewhat broad for counter rotation with mean widths of 7$\%$ and, respectively, 24$\%$($N=8192$). For this reason, we express the change of stability as trends $r^\pm_g$ in (\ref{E3}). The first, $r^+_g$($N=512$), is in excellent agreement with Dvorak (1986). For all $N$, the second evidences more chaotic change of stability. 

The ratio of the gap of corotation to that of counter rotation satisfies 
\begin{eqnarray}
r^+_g/r^-_g = 2.57 - 3.43e + 3.50e^2 -1.67e^3
\label{E6}
\end{eqnarray}
     (Fig.~\ref{Fig6}). The habitable zones (HZ) may be populated below the gap of corotation, outside the gap of counter rotation (Fig.~\ref{Fig3}). This may around low-luminosity binaries, where the HZ will be rather close in. In such a case, the HZ  may be exclusively populated by exoplanets on counter rotating orbits. This suggests it may be of interest to search for just such exoplanets around low-luminosity binaries in future observations, perhaps around WD-WD binaries.
In a future development, we plan to include ray-tracing to produce synthetic light curves of exoplanets with and without exomoons to assess potential detectability of selected systems from the existing exoplanet catalogues, see for instance, Zucker \& Giryes (2017). Studies of this kind provide detailed injection experiments to advanced detection algorithms to test and quantify a variety of statistical and system parameters. This may serve as input to advanced search and detection strategies with above mention upcoming advanced telescopes.


%

\section*{Acknowledgements}
Support is acknowledged from the National Research Foundation of Korea under grants 2015R1D1A1A01059793, 2016R1A5A1013277 and 2018044640.






\begin{thebibliography}{99}
\bibitem[\protect\citeauthoryear{Bashi et al.}{2018}]{DolevBashi2018}
Bashi, D., Helled, R., Zucker, S., 2018, Geosc., 8, 325 
\bibitem[\protect\citeauthoryear{Cuntz}{2015}]{Cuntz2015}
Cuntz, M., 2015, ApJ, 798, 101
\bibitem[\protect\citeauthoryear{Doyle et al}{2011}]{doy11}
Doyle, L. R., et al., 2011, Science, 333, 1602
\bibitem[\protect\citeauthoryear{Dvorak}{1986}]{Dvorak1986}
Dvorak, R., 1986, A\&A, 167, 379
\bibitem[\protect\citeauthoryear{Desidera \& Barbieri}{2007}]{des07}
Desidera, S. \& Barbieri, M., 2007, A\&A, 462, 345
\bibitem[\protect\citeauthoryear{Eggenberger et al.}{2004}]{egg04}
Eggenberger, A., et al., 2004, A\&A, 417, 353
\bibitem[\protect\citeauthoryear{Eggenberger et al.}{2007}]{egg07}
Eggenberger, A., et al., 2007, A\&A, 474, 273
\bibitem[\protect\citeauthoryear{Holman \& Wiegert}{1999}]{Holman1999}
Holman, M.J., Wiegert, P.A., 1999, AJ, 117, 621
\bibitem[\protect\citeauthoryear{Kempton et al.}{2018}]{kem18}
Kempton, E. M.-R., et al., 2018, PASP, 130, 993
\bibitem[\protect\citeauthoryear{Kasting et al.}{2013}]{kas93}
Kasting, J.F., et al., 1993, ApJ, 770, 82
\bibitem[\protect\citeauthoryear{Kopparapu et al.}{2013}]{kop13}
Kopparapu, R.K., Ramirez, R., 2013, ApJ, 765, 131
\bibitem[\protect\citeauthoryear{Landau \& Lifshitz}{1969}]{L.D.1969}
Landau, L.D., Lifshitz, E.M., {\it Mechanics} (Pergamon Press)
\bibitem[\protect\citeauthoryear{Lingam \& Loeb}{2017}]{Lingam2017}
Lingam, M., Loeb, A., 2017, PNAS, 114, 6689
\bibitem[\protect\citeauthoryear{Mayor \& Queloz}{1995}]{Mayor1995}
Mayor, M.,\& Queloz, D., 1995, Nature, 378, 355 
\bibitem[\protect\citeauthoryear{Morais \& Giuppone}{2012}]{Morais2012}
Morais, M.H.M., Giuppone, C.A., 2012, MNRAS, 424, 52
\bibitem[\protect\citeauthoryear{MathWorks Inc.}{2017}]{MathWork}
MATLAB 2017b, https://kr.mathworks.com/help/matlab/ref/ode45.html
\bibitem[\protect\citeauthoryear{Pilat-Lohinger et al.}{2002}]{Pilat2002}
Pilat-Lohinger, E., Funk, B., Dvorak, R., 2003, A\&A, 400, 1085 
\bibitem[\protect\citeauthoryear{Roell et al.}{2012}]{roe12}
Roell, T., et al., 2012, A\&A, 542, A92
\bibitem[\protect\citeauthoryear{Snellen}{2014}]{sne14}
Snellen, I.A.G., et al., 2014, Nature ,509, 63
\bibitem[\protect\citeauthoryear{van Putten}{2017}]{vanputten2017}
van Putten, M.H.P.M., 2017, NewA, 54, 115
\bibitem[\protect\citeauthoryear{Zeebe}{2017}]{Zeebe2017}
Zeebe R. E., 2017, ApJ, 154, 193

\end{thebibliography}



\label{lastpage}
\end{document}